\documentclass[12pt]{article}

\usepackage{graphicx}
\usepackage{amsmath,amssymb}
\usepackage{amsfonts}

\newcommand{\re}[1]{(\ref{#1})}
\newcommand{\ve}[1]{\mathbf{#1}}

\begin{document}
\begin{center}
{\bf  KINETICS OF PARTON- ANTIPARTON PLASMA
VACUUM CREATION IN THE TIME - DEPENDENT CHROMO - ELECTRIC FIELDS OF
ARBITRARY POLARIZATION} 

\vspace*{6mm}
{A.V. Filatov, S.A. Smolyansky, A.V. Tarakanov }\\      
{\small \it Physical Department of Saratov State University,
410026, Saratov, Russia   \\
E- mail: smol@sgu.ru   
            }
\end{center}

\vspace*{6mm}

\begin{abstract}
 The kinetic equation of non - Markovian type for description of the
vacuum  creation of parton - antiparton pairs under  action of a space homogeneous
time - dependent chromo - electric field of the arbitrary polarization is obtained on the strict
non - pertubative foundation in the framework of the oscillator representation. A comparison
of the effectiveness of vacuum creation with the case of linear polarization one is fulfilled.
\end{abstract}

\vspace*{6mm}

\section{ Introduction }

The Schwinger effect \cite{SHS}
of the vacuum production of electron-positron pairs (EPP's) under
the action of electromagnetic fields is one from a few QED
effects, that has not up to now an accurate experimental test. It
is stipulated by the huge electric fields $ E\sim E_{c} =
1,3\cdot10^{16} V/cm$ for the electrons that is necessary for
observation of this effect in a constant field. Such field strength
is unachievable for static fields
therefore main attention was involved the theoretical study of
pair creation by time-varying electric fields
(\cite{Brezin}-\cite{Marinov}).
The detailed description was obtained for the case of linearly
polarized spatially homogeneous time dependent electric field.
The sufficiently strong electric field can be achieved nowadays with
laser beams only. The estimations made before
(\cite{Brezin}-\cite{Popov})
showed that pair
creation by a single laser pulse with $E\ll E_c$ could be hardly
observed. The more optimistic results have been obtained for a planning
X-ray free electron lasers (\cite{Ringwald}-\cite{V02}) and for the
counter-propagating laser beams of optical range
(\cite{Av02}-\cite{20}).

In the present work we make  step on the way of
theoretical research of the parton - antiparton vacuum creation in the nonstationary
chromo - electric field of arbitrary polarization. The corresponding
kinetic equation (KE) will be derived below on the strict
non-perturbative dynamics basis. We will restrict ourself here by
consideration of the nonstationary Schwinger effect in vacuum only
leaving in a site the analysis of this effect in some
plasma-similar medium (see, e.g., \cite{Bulanov}). We use
the oscillator representation (OR) for the construction of the
kinetic theory (initially, this representation was be suggested in
the scalar QED \cite{Pervushin}). OR leads in the shortest way to
quasipartical (QP) representation, in which all dynamical
operators of observable quantities have diagonal form
\cite{Grib94}. On this basis, the Heisenberg-like equations of
motion for the creation and annihilation operators will be
obtained in the spinor QED (Sect.2) that corresponds to a large N in QCD.
 The corresponding kinetic
theory will be constructed in Sect. 3. The preliminary
communication about these results has made on the conference
\cite{Kiel04}. In general case, the obtained
KE of the non - Markovian type is rather
complicated because of spin effects. The case of rather weak of
the chromo - electric external
field is considered in the Sect. 4. The short conclusions are
summarized in Sect. 5.

We use the metric $g^{\mu\nu} =diag(1,-1,-1,-1)$ and the natural
units $\hbar = c =1$.

\section{ Oscillator representation}

Let us consider the QED system in the presence of an external
quasi-classical spatially homogeneous time-dependent electric
field of arbitrary polarization with the  4-potential (in the
Hamilton gange) $A^{\mu}(t)=(0,\ve{A(t)})$ and the corresponding
field strength $\ve{E}(t)=-\ve{\dot{A}}(t)$ (the overdots denote
the time derivative). Such a field can be considered either as an
external field, or as a result of the mean field approximation
\cite{Lif}. The Lagrange function is
\begin{equation}\label{1}
\mathcal{L} = \frac{i}{2} \{ \overline{\psi} \gamma^\mu D_{\mu}
\Psi - (D^{*}_{\mu} \overline{\psi}) \gamma^\mu \Psi\} -m
\overline{\psi} \psi,
\end{equation}
where $D_{\mu}=\partial_{\mu} + ieA_{\mu}(t)$ and -e is the
electron charge. The equations of motion are
\begin{eqnarray}\label{2}
(i \gamma^\mu D_\mu -m)\psi =0, \nonumber \\
\overline{\psi} (i \gamma^\mu \overleftarrow{D}_\mu^* +m)=0,
\end{eqnarray}
where $\bar{\psi} = \psi^+ \gamma^0$. The fields $\psi$ and
$\psi^+$ compose the pair of canonical conjugated variables. The
corresponding Hamiltonian is (k=1,2,3)
\begin{equation}\label{3}
H(t) = i \int d^3 x \psi^+ \dot{\psi} = \int d^3 x \bar{\psi} \{
-i \gamma^k D_k + m \}\psi.
\end{equation}

In the considered case, the system is space homogeneous and
nonstationary. Therefore the transition in the Fock space can be
realized on the basis functions $\phi= \exp{(\pm i\ve{k}\ve{x})}$
and creation and annihilation operators become the time dependent
one, generally speaking. Hence, we have the following
decompositions of the field functions  in the discrete momentum
space ($ V = L^3$ and $p_i = (2\pi/L )n_i$ with an integer $n_i$
for each $i=1,2,3$):
\begin{eqnarray}\label{4}
\psi (x) &=& \frac{1}{\sqrt{V}} \sum_{\ve{k}} {\sum_{\alpha=1,2}}
\left\{ e^{i\ve{k}\ve{x}} a_{\alpha}(\ve{k},t)
u_{\alpha}(\ve{k},t) + e^{-i \ve{k}\ve{x}} b_{\alpha}^+ (\ve{k},t)
v_{\alpha}(\ve{k},t) \right\}, \nonumber \\
\bar{\psi}(x) &=& \frac {1}{\sqrt{V}}\sum_{\ve{k}} \sum_{\alpha
=1,2} \{ e^{-i\ve{k}\ve{x}}a^+ _{\alpha} (\ve{k},t)
\bar{u}_{\alpha}(\ve{k},t) + e^{i\ve{k}\ve{x}} b_{\alpha}
(\ve{k},t)\bar{v}_{\alpha} (\ve{k},t)\}.
\end{eqnarray}

The nearest aim is derivation of the equations of motion for
the creation and annihilation operators on the basis of the
primary equations \re{2} and the use of the free $u,v$-spinors as
the basic functions with the natural substitution of the canonical
momentum with the corresponding kinematic one (that corresponds to the basic OR idea). It is necessary to
take into account, that the electron and positron states are
different by sings of the charges and hence their kinematic
momentum are $\ve{p}=\ve{k}-e\ve{A}$ for electrons and
$\ve{p}^c=\ve{k}+e\ve{A}$ for positrons. Thus, the following
"free-like" equations for the spinors are postulated in OR:
\begin{eqnarray}\label{5}
[{\gamma}p - m]u(\ve{k},t)=0, \nonumber
\\ {[{\gamma}p^c + m]} v(\ve{k},t)=0,
\end{eqnarray}
where $p^0 =\omega(\ve{p})=\sqrt{m^2 +\ve{p}^2}$. These equations
have the orthogonal solutions which is convenient to normalize on
unit \cite{Greiner,Bogol}
\begin{align}\label{6}
{{u}}^+_{\alpha} (\ve{k}, t) {v_{\beta}} (-\ve{k},t) = 0&,
\nonumber \\
u^+_{\alpha}(\ve{k},t) u_{\beta}(\ve{k},t) =
v^+_{\alpha}(-\ve{k},t) v_{\beta}(-\ve{k},t) =
\delta_{\alpha\beta}&, \nonumber\\ {\bar{u}}_{\alpha}(\ve{k},t)
u_{\beta}(\ve{k},t) = \frac{m}{\omega
(\ve{k},t)}{\delta}_{\alpha\beta}, \qquad
{\bar{v}}_{\alpha}(\ve{k},t) v_{\beta}(\ve{k},t) =
-\frac{m}{\omega (\ve{k},t)}{\delta}_{\alpha\beta}&,
\end{align}

The decompositions \re{4} and the relation \re{6} lead to the
diagonal form of the Hamiltonian \re{3} at once (before second
quantization)
\begin{equation}\label{7}
H(t) = \sum_{\ve{k}, \alpha}\omega(\ve{k},t)\left[a_{\alpha}^+
(\ve{k},t) a_{\alpha}(\ve{k},t) - b_{\alpha}(-\ve{k},t)
b^+_{\alpha} (-\ve{k},t)\right].
\end{equation}
Such form of the Hamiltonian is necessary for
interpretation of the time dependent operators $a^+ ,a$
(and $b^+ ,b$) as the operators of creation and
annihilation of quasi-particles (anti-quasi-particles).
Thus, this way results to QP representation at once.

Now, in order to get the equations of motion for the creation and
annihilation operators in the OR, let us substitute the
decomposition \re{4} in the Eq.\re{2} and use the relations
\re{6}. Then we obtain as the intermediate result the following
closed system of equations of motion in the matrix form:
\begin{align}\label{8}
\dot{a}(\ve{k},t) + U_{(1)}(\ve{k},t)a(\ve{k},t) +
U_{(2)}(\ve{k},t)b^+(-\ve{k},t) &= -i \omega
(\ve{k},t)a(\ve{k},t), \nonumber\\ \dot{b}(-\ve{k},t) -
b(-\ve{k},t)V_{(2)}(\ve{k},t) + a^+(\ve{k},t) V^+_{(1)}(\ve{k},t)
&= - i\omega (\ve{k},t)b(-\ve{k},t).
\end{align}
The spinor constructions was introduced here
\begin{align}\label{9}
U_{(1)}^{\alpha\beta}(\ve{k},t) &= {u}^+_\alpha
(\ve{k},t)\dot{u}_{\beta}(\ve{k},t),& U^+_{(1)} &= - U_{(1)},
\nonumber \\ U_{(2)}^{\alpha\beta)}(\ve{k},t) &=
{u}^+_{\alpha}(\ve{k},t)\dot{v}_{\beta}(-\ve{k},t),&
U^+_{(2)} &= - V_{(1)}, \nonumber \\
V_{(2)}^{\alpha\beta}(\ve{k},t) &=  {v}^+_{\alpha}
(-\ve{k},t)\dot{v}_{\beta}(-\ve{k},t),&  V^+_{(2)} &= - V_{(2)}.
\end{align}
The matrices $U_{(2)}$ and $V_{(1)}$ describe transitions between
states with the positive and negative energies and different spin
while the matrixes $U_{(1)}$ and $V_{(2)}$ show the spin rotations
in the external field $\ve{A}^k(t)$.

The equations \re{8} are compatible with the standard
anti-commutation relations because the matrix $U_{(1)}$ is
anti-hermitian:
\begin{equation}\label{10}
\{a_{\alpha}(\ve{k},t), a^+_{\beta}(\ve{k}', t)\} =
\{b_{\alpha}(\ve{k},t), b^+_{\beta}(\ve{k}', t)\} =
\delta_{\ve{k}\ve{k}'}\delta_{\alpha\beta}.
\end{equation}

Let us write the $u,v$-spinors in the explicit form using the
corresponding free spinors \cite{Raider}:
\begin{align}\label{13}
u^+_1(\ve{k},t) = A(\ve{p}) \begin{bmatrix} \omega_+ , 0 , p^3
,p_- \end{bmatrix}, && u^+_2(\ve{k},t) = A(\ve{p})
\begin{bmatrix} 0, \omega_+ ,  p_+
, -p^3 \end{bmatrix}, \nonumber \\
v^+_1(-\ve{k},t) = A(\ve{p}) \begin{bmatrix}  -p^3, - p_- ,
\omega_+ , 0 \end{bmatrix}, && v^+_2(-\ve{k},t) = A(\ve{p})
\begin{bmatrix} -p_+ , p^3 ,0, \omega_+
\end{bmatrix},
\end{align}
where $p_{\pm} = p^1 \pm i p^2$, $\omega_+ =\omega +m$ and
$A(\ve{p}) = [2\omega\omega_+]^{-1/2}$. In this representation
$U_{(1)} = V_{(2)}$ and $U_{(2)} = - V_{(1)}$ so a sufficient set
is
\begin{align}\label{14}
U_{(1)}(\ve{k},t) & = i \omega a [\ve{p}\ve{E}]\ve{{\boldsymbol\sigma}},&
U_{(2)}(\ve{k},t) & = \ve{q}\ve{\boldsymbol\sigma},
\end{align}
where $\sigma^k$ are the Pauli matrices, $\ve{q} = a [\ve{p}
(\ve{p}\ve{E}) -\ve{E} \omega \omega_+ ]$ and
$a=e/2\omega^2\omega_+$.

The operator equations of motion \re{8} become more simple:
\begin{align} \label{16}
\dot{a}(\ve{k},t) &=  - U_{(1)}(\ve{k},t)a(\ve{k},t)-U_{(2)}
b^+(-\ve{k},t) -i\omega (\ve{k},t)a(\ve{k},t), \nonumber\\
\dot{b}(-\ve{k},t) &=   b(-\ve{k},t) U_{(1)}(\ve{k},t) +
a^+(\ve{k},t)U_{(2)}(\ve{k},t) - i\omega (\ve{k},t)b(-\ve{k},t).
\end{align}

\section{Kinetic equation (the general case)}

In order to get KE for time dependent electric fields of arbitrary
polarization, let us introduce the one particle correlation
functions of electrons and positrons
\begin{align} \label{19}
f_{\alpha\beta}(\ve{k},t) &=\, < a^+_{\beta}(\ve{k},t)
a_{\alpha}(\ve{k},t)>, \nonumber \\
{f}^c_{\alpha\beta}(\ve{k},t) &=\, < b_{\beta}(-\ve{k},t)
b^+_{\alpha}(-\ve{k},t)>,
\end{align}
where the averaging procedure is performed over the in-vacuum
state \cite{Grib94}. The diagonal parts of these correlators are
connected with relations
\begin{equation}\label{19a}
\sum\limits_{\ve{k},\alpha} \bigl( f_{\alpha\alpha}(\ve{k},t) +
f^c_{\alpha\alpha}(\ve{k},t) \bigr) = Q,
\end{equation}
where $Q$ - total electric charge of the system. Differentiation
over time leads to equations
\begin{align}\label{20}
\dot{f} &= [f,U_{(1)}] - \bigl( U_{(2)} f^{(+)} + f^{(-)}U_{(2)}
\bigr) ,\nonumber \\ \dot{f}^c &= [f^c,U_{(1)}] + \bigl(
f^{(+)}U_{(2)} + U_{(2)} f^{(-)}\bigr) ,
\end{align}
where the auxiliary correlation functions was introduced
\begin{align}\label{21}
f^{(+)}_{\alpha\beta}(\ve{k},t) & = \,<a^{+}_{\beta}(\ve{k},t)
b^{+}_{\alpha}(-\ve{k},t)>, \nonumber \\
f_{\alpha\beta}^{(-)}(\ve{k},t) & =\,
<b_{\beta}(-\ve{k},t)a_{\alpha}(\ve{k},t)>.
\end{align}
The equations of motion for these functions can be obtained
similarly:
\begin{align}
\dot{f}^{(+)} =  [ {f}^{(+)} , U_{(1)} ] + \bigl( U_{(2)} f - f^c
U_{(2)}
\bigr) + 2i\omega f^{(+)}, \nonumber \\
\dot{f}^{(-)} = [ {f}^{(-)} , U_{(1)} ] + \bigl( f U_{(2)} -
U_{(2)} f^c \bigr) - 2i\omega f^{(-)} \label{22}
\end{align}
with the connection $\stackrel{+}{ f^{(+)}} = f^{(-)}.$
In general case, the Eqs.\re{20} and\re{22} represent the closed
system of 16 ordinary differential equations.

Accounting of charge symmetry (in consequence of that $f^c=1-f$
allows to reduce this number up to 12. If to express the anomalous
correlators \re{21} via the original functions \re{19} with help
of Eqs.\re{20}, it can obtain the closed KE in the
integro-differential form \cite{Kiel04}. Let us write this KE of
non-Markovian type in the following matrix form:
\begin{multline}\label{23}
\dot{f}(t) = [f(t), U_{(1)}] - U_{(2)}(t) S(t) \int\limits_{t_0}^t
dt' S^{+}(t') [U_{(2)}(t') f(t') - {f}^c(t')U_{(2)}(t')] S(t')
S^+(t')e^{2i \theta (t,t')} \\ - S(t) \int\limits_{t_0}^t dt'
S^{+}(t') [f(t') U_{(2)}(t')  - U_{(2)}(t'){f}^c(t') ] S(t')
S^+(t') U_{(2)}(t) e^{-2i \theta (t,t')},
\end{multline}
where the evolution operator of the spin rotations $S(\ve{k},t)$
is defined by equation
\begin{equation}\label{24}
\dot{S} = - U_{(1)}(t) S(t)
\end{equation}
with the initial condition $S(t_0)=1$ ($t_0$ is some initial time)
and $\theta(t,t')=\theta(t)-\theta(t')$,
\begin{equation}\label{25}
\theta(t)= \int\limits_{t_0}^t dt' \omega(\ve{k},t').
\end{equation}

In comparison with the KE for the known case of the linear
polarized field
\begin{equation}\label{26}
\ve{A}(t) = \{0,0,A^3(t)=A(t)\},
\end{equation}
KE \re{23} has more complicated form because  nontrivial spin
effects. In general case, KE \re{23} is not allow simplification
because of $[U_{(1)}, U_{(2)}] \neq 0$.

\section{Perturbation theory}

Let us write the source term (the right hand side) of KE \re{23}
in the leading (second) order of the perturbative theory with
respect to weak external field, $E_{m}/E_{c}\ll 1$. The adiabatic
parameter \cite{Popov} $
\gamma = \frac{m\nu}{eE_m} $ is arbitrary (here
$E_m$ is amplitude of external electric field, $\nu$ is it
characteristic frequency). In according to the relations \re{14},
$U_{(1)} \sim U_{(2)} \sim E_m $ in the leading approximation.
 Then in the leading order it is necessary to put
$S\to S_0 =1$ according to Eq.\re{24}.

We take into account also electroneutrality of the system and
relation \re{10}, so $f^c=1-f$. In the considered leading
approximation, the diagonal terms of the correlation functions
\re{14} if small in comparison with unit, $f_{\alpha\alpha}$, and
the non-diagonal terms $f_{\alpha\beta} \sim E^2 $ for $\alpha
\neq \beta$, that allows to omit the corresponding contribution in
the source term
\begin{equation}\label{29}
\dot{f}(t) = \int\limits_{t_0}^t Sp \{ U_{(2)}(t)
U_{(2)}(t')\} \cos{2\theta(t,t')}.
\end{equation}
As it follows from Eq. \re{14} ($\omega_{+}=\omega_{0}$),
\begin{align}\label{30}
2 Sp \{ U_{(2)}(t) U_{(2)}(t')\} = \frac{e^2}{2\omega^{2}\omega_0^2}
\left\{ \ve{E}(t)\ve{E}(t') \omega\omega_0 -
(\ve{E}(t)\ve{p})(\ve{E}(t')\ve{p})\right\} = \Phi(\ve{p}|t,t').
\end{align}
If at the initial time before switch-on of an electric field the
electrons and positrons are absent, we can write the total density
of quasiparticles
\begin{equation}\label{31}
n(t) = \frac{1}{4\pi^3} \int d^3p \int\limits_{t_0}^t dt_1
\int\limits_{t_0}^{t_1} dt_2 \Phi(\ve{p}|t_1,t_2)
\cos{[2\theta(t_1,t_2]}.
\end{equation}

In the case of the linear polarization \re{26}, from Eqs. \re{30}
and \re{31} it follows the well known result \cite{Tar,20}:
\begin{equation}\label{32}
n(t) = \frac{1}{4\pi^3} \int d^3p \left| \int\limits_{t_0}^t dt'
\lambda(t') \exp{(2i\theta(t,t'))} \right|^2 ,
\end{equation}
where $\lambda(\ve{p},t) = eE(t) \varepsilon_\perp / 2\omega^2$
and $\varepsilon_\perp^2 = m_2+p_\perp^2$, $\ve{p}_\perp$ is the
transversal momentum relatively of the vector $\ve{E}(t)$.

The relations \re{31} and \re{32} are convenient for the numerical
analysis, that is planned to made in the following work.

\section{ Conclusion}

Thus, it was shown that the oscillator representation may be used
 for the KE derivation in the rather non-trivial
case of the time-dependent chromo-electric field of arbitrary
polarization. The obtained KE's can be used for investigation of
particle-antiparticle vacuum creation in strong laser fields of optical and X-ray range
as well as in the chromo-electric fields acting in the pre-equilibrium stage of QGP evolution.
Besides, the used method opens prospects for further
generalization ( e.g., the account of a constant magnetic field).


\end{document}